\documentclass[journal]{IEEEtran}
\usepackage{cite}
\usepackage{balance}
\usepackage{verbatim}
\usepackage{algorithmic}
\usepackage{graphicx}
\usepackage{graphics}
\usepackage{textcomp}
\usepackage{float}
\usepackage{xcolor}
\usepackage{comment}
\usepackage[linesnumbered,ruled,vlined]{algorithm2e}
\usepackage{fullpage}
\usepackage{times}

\usepackage{amsmath,amssymb,amsfonts}
\usepackage{algorithmic}
\graphicspath{{./}}
\usepackage[section]{placeins}
\usepackage{textcomp}
\usepackage{booktabs}
\usepackage{tabularx}
\usepackage{multirow}
\usepackage{multicol}
\usepackage{dcolumn}
\usepackage{booktabs}
\usepackage{soul}
\usepackage{color}

\usepackage{enumerate}

\usepackage{float}

\usepackage{caption}
\usepackage{subcaption}

\def\BibTeX{{\rm B\kern-.05em{\sc i\kern-.025em b}\kern-.08em
    T\kern-.1667em\lower.7ex\hbox{E}\kern-.125emX}}
\begin{document}

\author{Zakria Qadir, ~\IEEEmembership{Member,~IEEE,} Muhammad Bilal, ~\IEEEmembership{Senior Member,~IEEE,} Guoqiang Liu and Xiaolong Xu, ~\IEEEmembership{Senior Member,~IEEE}

\thanks{Z. Qadir is with the AI Institute, University of New South Wales, Australia. Email: z.qadir@unsw.edu.au.}

\thanks{M. Bilal is with the School of Computing and Communications, Lancaster University, UK. Email:
m.bilal@ieee.org.}

\thanks{Guoqiang Liu is with the School of Software, Nanjing University of Information Science and Technology, China. Email: 202212210015@nuist.edu.cn.}

\thanks{Xiaolong Xu is with the School of Software, Jiangsu Province Engineering
Research Center of Advanced Computing and Intelligent Services, Jiangsu Collaborative Innovation Center of Atmospheric
Environment and Equipment Technology (CICAEET), Nanjing University of Information Science and Technology, China. Email: xlxu@ieee.org.}
}

\title{Autonomous Trajectory Optimization for UAVs in Disaster Zone Using Henry Gas Optimization Scheme}

\maketitle

\begin{abstract}

The unmanned aerial vehicles (UAVs) in a disaster-prone environment plays important role in assisting the rescue services and providing the internet connectivity with the outside world. However, in such a complex environment the selection of optimum trajectory of UAVs is of utmost importance. UAV trajectory optimization deals with finding the shortest path in the minimal possible time. In this paper, a cluster optimization scheme (COS) is proposed using the Henry gas optimization (HGO) metaheuristic algorithm to identify the shortest path having minimal transportation cost and algorithm complexity.  The mathematical model is designed for COS using the HGO algorithm and compared with the state-of-the-art metaheuristic algorithms such as particle swarm optimization (PSO), grey wolf optimization (GWO), cuckoo search algorithm (CSA) and barnacles mating optimizer (BMO). In order to prove the robustness of the proposed model, four different scenarios are evaluated that includes ambient environment, constrict environment, tangled environment, and complex environment. In all the aforementioned scenarios, the HGO algorithm outperforms the existing algorithms. Particularly, in the ambient environment, the HGO algorithm achieves a 39.3\% reduction in transportation cost and a 16.8\% reduction in computational time as compared to the PSO algorithm. Hence, the HGO algorithm can be used for autonomous trajectory optimization of UAVs in smart cities.

\end{abstract}


%

\section{Introduction}
UAVs (unmanned aerial vehicles) or drones have gained rapid popularity over the last decade due to their applications in the context of smart cities. Multiple optimization techniques can be incorporated with UAVs to perform exceptional in emergency situations like bushfire disaster. Additionally, optimizing UAV trajectory for evacuating survivors in catastrophic situations is quite challenging \cite{43,44}. Specifically, the complex terrain and adverse weather conditions often present in disaster zones pose significant risks to UAV flight safety and stability \cite{mohsan2023unmanned}. Also, the sudden and unpredictable nature of disasters can damage communication infrastructure, leading to potential interruptions in communication between UAVs and ground control centers, thereby impacting command and control operations \cite{pereira2021optimal}. Furthermore, the instability of power supplies and limited energy reserves necessitate optimal strategies for energy consumption during path planning \cite{hong2021energy}. Finally, the urgency of rescue operations in disaster areas requires efficient and effective UAV deployment, making trajectory optimization critical for timely and successful mission execution \cite{damavsevivcius2023sensors}.

Optimization problem are referred to finding the best possible cost using the specified parameters keeping minimal algorithm complexity possible \cite{54}. The problems can be identified in any field of science, leading researchers towards developing complex and optimized algorithms to solve challenging problems \cite{55}. Contrarily, conventional optimization algorithms have certain shortcomings such as, local optima convergence, single based solutions and undefined search space. Subsequently, metaheuristic algorithms are designed to solve these complex and unsolved optimization problems, resulting in best global cost in minimal timeframe. 
In this technological era, Metaheuristics have gained considerable popularity in various optimization problems. Metaheuristic algorithms leverages path planning optimization for UAVs in any disaster based scenario such as, (1) adaptability \cite{45}, (2) local optima avoidance\cite{46} and (3) no requirement of multivariable generalization make these metaheuristic algorithm far better than conventional algorithms \cite{47}.

Metaheuristic algorithms are of two types; single based and population based. Regardless of its nature, most of the metaheuristic algorithms comprise of population based algorithm that have two important phases; Exploration and Exploitation phase \cite{50}. All Metaheuristic algorithm achieves good efficiency by balancing between these two exploration and exploitation phases in the search space. Moreover, exploration phase generates new solutions to avoid local optima, however,   exploitation phase generates similar solutions to improve the previously generated solution leading towards efficient convergence \cite{51}.

Different Metaheuristic algorithms have been incorporated for different UAV path planning but there remains a major research question whether these algorithms are enough for solving all the critical problems \cite{52}. The answer to this question is No Free Lunch (NFL), as it states that all the optimization problems cannot be solved by only one metaheuristic algorithm, we have to hybridize or use new algorithm for every different problem \cite{53}.

\begin{figure}[t]
\center
\includegraphics[width=3.2in, height = 6.5cm]{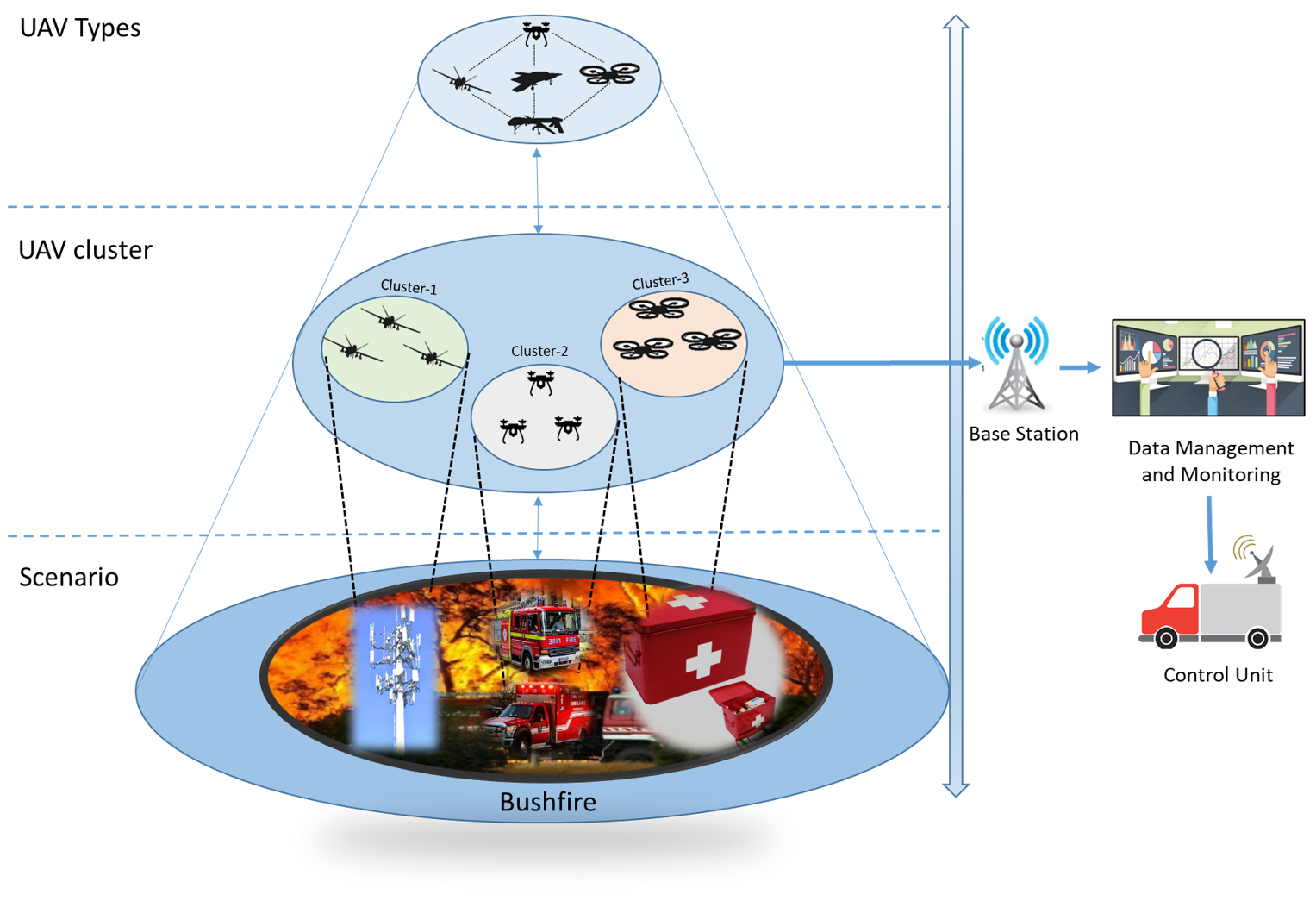} 
\caption{3D COS for UAVs in a disaster scenario}
\label{Figure1}
\vspace{-4 mm}
\end{figure}

In this paper, we examine a scenario involving a bushfire disaster and categorize various UAV types into distinct clusters, each assigned specific tasks within a proposed timeframe. Figure \ref{Figure1} illustrates three clusters incorporated into the system. Cluster 1 is designated to monitor and ensure seamless communication between UAVs and the ground station, which guarantees efficient communication in the sudden disaster. Cluster 2 is responsible for providing live streaming of the bushfire-affected area, updating the rescue team about survivors, and informing neighboring UAVs. Cluster 3 is tasked with supplying logistics to the rescue team, including first aid and other medical equipment for the initial treatment of survivors. As a result, cluster 2 and cluster 3 achieve efficient and effective UAV deployment in during urgent disaster relief.

It should be noted that this paper involves the experimental simulation of a disaster environment as creating a real-world disaster environment will be quite difficult. To further strengthen our point, we have included some papers that have used the experimental simulation environment \cite{40, 41, 42}.
Besides, many researchers have proposed novel solutions for UAV trajectory optimization in order to reduce energy consumption and improve rescue efficiency, still, it is hard to efficiently optimize the trajectory comprising of many obstacles and COS in a dynamic environment.

\subsection{Main Contribution}
In this paper, a COS is proposed based on HGO metaheuristic-based optimization algorithm and compared with state-of-the-art optimization algorithms such as BMO, CSA, GWO, and PSO. The main contribution of this paper is as follows: 
\begin{itemize}
\item COS is incorporated to differentiate between different clusters of UAVs in order to monitor the affected area and provide timely feedback. 
\item HGO algorithm shows significant improvement in the transportation cost and computation time compared with other state-of-the-art metaheuristic algorithms. 
\item The efficient ability of the HGO algorithm to balance the exploration and exploitation phases within the COS makes the optimization more robust. 
\item Subsequently, the local optima are reduced for complex routing scenarios and helps in finding the best global solution in the minimal possible time.
\end{itemize}

\subsection{Paper Notations and Organization} 
The organization of the paper is as follows: Section I highlights the main contribution of the paper. The related study for metaheuristic algorithm is presented in Section II. Section III discusses the mathematical modeling along with pseudo code for HGO algorithm having Exploration and Exploitation phase making optimization more robust. Section IV defines the workspace, problem statement and the trajectory optimization for post bushfire assessment. The experimental results for four different cases are evaluated  based on trajectory optimization in Section V. Finally, we conclude the paper in Section VI. The paper notations are listed in Table \ref{table:notations}.
\FloatBarrier
\begin{table}[!htbp]
\centering
\caption{List of Defined Symbols}
\label{table:notations}
\resizebox{\columnwidth}{!}{%
\begin{tabular}{@{}cc@{}}
\toprule
Symbol & Abbreviation \\ \midrule
$U_p$ & Position of UAV \\
$U_{max}, U_{min}$ & Upper and lower bounds \\
$U_s$ & Start point of UAV \\
$U_d$ & Destination of UAV \\
$G_{p,q}$ & Position of worst UAV \\
$G_{max}, G_{min}$ & Upper and lower bounds \\
$c_1, c_2, c_3$ & Control points \\
$H_q$ & Henry Constant \\
$P_{p,q}$ & UAV velocity \\
$X_{best}$ & Fitness of best UAV \\
$X_{p,q}$ & Fitness of UAV in cluster \\
$F$ & Flag (diversity) \\
$C_q, O_1=1, O_2=10, O_3=1, K^{\theta}=298.15, \varepsilon=0.05, \omega$ & Constants \\
$rand$ & Random number between 0 and 1 \\
$max_{iter}$ & Maximum iteration \\
$iter$ & Iteration \\
$T_{p,q}$ & Updated position of UAV \\
$\alpha$ & Interaction ability of UAV \\
$\beta$ & Influencing ability of UAV \\
\bottomrule
\end{tabular}%
}
\end{table}

\section{Related Study}
Metaheuristic algorithms have no unique criteria for classification, rather they have various sources of inspiration. Accordingly, they can be categories into three main categories as follows: (1) natural science based algorithms (NSAs), (2) nature-inspired algorithm (NIA) consist of bioinspired algorithms (BIA) and swarm intelligent based algorithm (SIA) and (3) natural phenomena based algorithm (NPA) \cite{meta1}.

The first category: NSA mimics the physical law and chemical reaction e.g., pressure, gravity, particle movement, etc. Most popular algorithms are Gravitational Search Algorithm (GSA), Search Annealing (SA) and HGO. HGO algorithm is proposed in this paper that is evaluated based on the Henry's gas law \cite{hgo2}.

The second category: NIA algorithm mimics the behavior of birds, fishes, ants, and also perform biological evolution \cite{meta2}. The main source of inspiration is their search for food, following other individuals and flocking. A random population is generated at the start to begin the search process and the process ends when it reaches the global best solution, or the maximum number of iterations is reached. Some popular NIA algorithms consist of PSO, GWO, CSA, BMO, Ant Colony Optimization (ACO), GA and others.

The third category: NPA mimics the emotional, social, and natural phenomena from different sources. Most popular algorithm consist of Virus Colony Search (VCS) and Backtracking Optimization Search (BOS) \cite{meta3}. 
 
As aforementioned, HGO is NSA based algorithm and is compared with another class of NIA algorithms that includes PSO\cite{psom,pso4}, GWO\cite{gwom,gwo3}, CSA\cite{csam,csa2} and BMO\cite{bmom}. HGO combines gas behavior simulation, global search mechanisms, and dynamic adaptation. By simulating the movement, collisions, and diffusion of gas particles, it effectively explores the solution space, avoids local optima, and enhances global search capabilities. Additionally, HGO algorithm has a unique feature of balancing the exploration and exploitation phases that make optimization robust and helps in finding the global best cost in minimal time span. In contrast, traditional BMO, CSA, GWO and PSO algorithms often lack flexibility and adaptability when dealing with high-dimensional problems, making it difficult to achieve accurate path planning in dynamic scenarios. PSO algorithm does not settle at global minima and produce random oscillations due to random number embedded in velocity vector \cite{pso,pso1,pso2,pso3}. GWO algorithm causes premature settling of particles, thus causes less explorative behavior over iterations \cite{gwo,gwo1,gwo2}. In CSA, the random oscillations caused due to Levy flight in exploitation phase can detract from global optima, thus causes instability in particle movement \cite{csa,csa1}. However, the premature convergence in BMO algorithm causes the overall accuracy to be limited \cite{bmo1,bmo}. 

In \cite{peng2019hybrid} authors presented a special hybrid genetic algorithm method that allows the truck and several UAVs to work together. The vehicle's routing and scheduling method lets numerous UAVs to carry various items to clients in different areas at the same time.

Shao et al.,\cite{9099809} proposed a novel long range service delivery drone using Ant Colony Optimization (ACO) algorithm. The authors uses multiple depots that reduces the path length between start and destination but overall increases the number of iterations and time complexity for the model. Rigas et al.,\cite{rigas2021scheduling} addresses the extension of Multiple Traveling Salesman Problem (MTSP) by incorporating greedy algorithm and ACO. Monitoring task is performed using fleet of drones having limited mobility and battery life. The experimental studies are conducted that depict UAV mobility in real world environment. Zhang et al,\cite{zhang2021data} proposes a forest fire monitoring mechanism using bi-level hybridization-based metaheuristic algorithm (BLHMA). The task of UAV is to hover the defined locations and collect environmental data from detection nodes that are deployed in high-risk forest area.

The first phrase in the two-stage procedure is always referred to as the Multiple Traveling Salesman Problem (MTSP) extension. The information collecting problem is defined as a version of MTSP in \cite{20}, and it entails arranging the search of a transportation route in the shortest time feasible. The first step is similarly defined by Stump et al., \cite{21} as the discovery of discrete site travel sequences, which is being solved using operational research methods. In \cite{22}, employed the MTSP model to identify the visiting pattern of intended site while avoiding banned regions and arriving at the target.
Other precise approaches, like as Depth-First Search (DFS) \cite{23}, Breadth-First Search (BFS) \cite{24}, and Branch and Bound (BB) \cite{25}, have been employed to identify systematically and objectively, even if they weren't the best, in an acceptable amount of time. Additionally, these methods are frequently unable to handle complex graphs with a significant number of variables or constraints. Artificial intelligence approaches \cite{26} and software-in-the-loop simulations \cite{27}, on the other hand, are frequently utilized due to their easier operation and reduced complexity \cite{28,29,30}.
ACO and SA are two different optimization strategies used in [9] to identify the shortest feasible path among various nodes. The GA is used in \cite{31,32} to swiftly determine a viable and quasi-optimal path while the mission's destinations are modified. Also published in \cite{33}, the Network-based Heterogeneous Particle Swarm Optimization (NHPSO) surpassed previous PSO algorithms on their test scenarios. A unique evolutionary search algorithm is used in \cite{34} and \cite{35} to collect organized information from previous solutions, which can then be employed to schedule pathways whenever a comparable issue arises. Research in these studies indicated that metaheuristic approaches are faster than precise algorithms at finding effective solution, especially when the number of variables is large \cite{36}.

\section{Mathematical Modeling of HGO}

\begin{figure}[h!]
\center
\includegraphics[width=3in, height = 3cm]{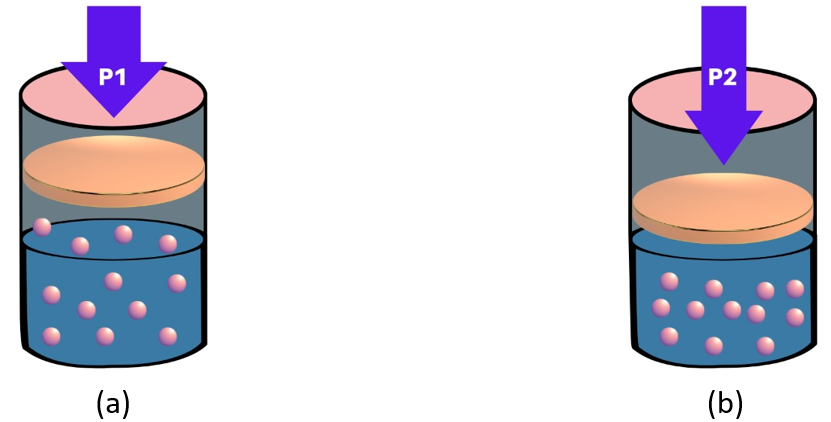} 
\vspace{2 mm}
\caption{Henry's Law for solubility of gas (a) solubility of gas particles in equilibrium at P1 (b) higher pressure P2 increases the solubility of gas}
\label{Figure2}
\vspace{-4 mm}
\end{figure}

In this section, HGO algorithm inspired by Henry’s law is presented. In 1803, William Henry formulated Henry law that states: The solubility of a gas in a given liquid is directly proportional to the amount of partial pressure in equilibrium state at constant temperature. The main working mechanism of Henry law can be seen in Figure \ref{Figure2} and can be equated as follows:

\begin{equation}
T\_{p,q}=H\_q \space P\_{p,q}
\end{equation}
where $T_{p,q}$ is defined as the solubility of gas, $H_q$ is the Henry constant and $P_{p,q}$ is the partial pressure inserted on the specified gas.
In this research, the main inspiration of using HGO metaheuristic algorithm is in the context of trajectory optimization of UAV swarm. Therefore, we referred the gas particles as swarm of UAVs, pressure is treated as the velocity of these UAVs and the solubility is treated as the trajectory optimization from starting point towards destination.

\FloatBarrier

The step wise execution of mathematical equations for the HGO algorithm are reported as follows:\\

\textbf{\emph{Step 1: Initialization.}}
The population of UAV swarm and their respective position are initialized as follows:

\begin{equation}
U\_{p}(t+1)=U\_{\min }+r (U\_{\max }-U\_{\min })
\end{equation}
where $U_{p}$ refers to the position of UAV, $U_{\max}$ \text{and} $U_{\min}$ are the upper and lower bounds, r indicates the random number (0-1) and t is the iteration time. The Henry constant $H_{q}(t)$, UAV velocity $P_{p, q}$ and constant $C_{q}$ are initialized using equation:

\begin{equation}
\begin{aligned}
&H\_{q}(t)=O\_{1}  \operatorname{rand}(0,1), P\_{p, q}=O\_{2}  \operatorname{rand}(0,1), \\
& C\_{q}=O\_{3}  \operatorname{rand}(0,1)
\end{aligned}
\end{equation}
where $O_{1}, O_{2}, O_{3}$ are the constant value equal to 1, 10, 1, respectively.\\

\textbf{\emph{Step 2: Cluster Analysis.}}
The population of UAV is divided into set of small UAV swarm called clusters, where every cluster has same Henry coefficient $H_{q}$.\\

\textbf{\emph{Step 3: Best Cost Evaluation.}}
Every cluster q identifies the best UAV based on the best global cost and rank the UAVs to obtain the optimal trajectory among the entire swarm. \\

\textbf{\emph{Step 4: Update Henry coefficient, optimizing UAV path and worst case UAV.}}
In this step we will first update the Henry coefficient using Equation \ref{eq3}, than update the position of optimized path for UAV using Equation \ref{eq4} and lastly will escape from local optimum by balancing the exploration and exploitation phase and rank the worse case UAVs and update their positions using Equation \ref{eq5}, \ref{eq6}, \ref{eq7}. 

\begin{equation}\label{eq3}
\begin{aligned}
&H\_{q}(t+1)=H\_{q}(t)  \exp (-C\_{q} (1 / K(t)-1 / K^{\theta}))\\
\end{aligned}
\end{equation}
where $K(t)=\exp (-t / i t e r), K^{\theta}$ is constant and equal to 298.15, $K(t)$ depends on $t$ (iteration time) and $iter$ (total number of iteration), respectively.
\begin{equation}\label{eq4}
T\_{\text{p}, q}(t)=\text{Const}  H\_{q}(t+1)  P\_{\text{p}, q}(t)
\end{equation}
where, $T_{\text{p}, q}$ refers to the optimized trajectory and $P_{\text{p}, q}$ is the UAV velocity and $Const$ as indicated by name is a constant value.

\begin{equation}\label{eq5}
\begin{aligned}
&U\_{p, q}(t+1)=U\_{p, q}(t)+F (rand ) \alpha (U\_{p, \text { best }}(t)\\
&-U\_{p, q}(t)) \\
&\quad+F ( rand ) \beta (T\_{p, q}(t)  U\_{\text {best }}(t)-U\_{p, q}(t)) \\
&\alpha=\omega  \exp (-\frac{X\_{\text {best }}(t)+\varepsilon}{X\_{p, q}(t)+\varepsilon}), \varepsilon=0.05
\end{aligned}
\end{equation}
where the UAV position is denoted by $U_{p, q}$, rand is the random constant. $U_{p, \text { best }}$ is the best UAV in cluster q, whereas, $U_{\text {best }}$ indicates the best UAV in the entire swarm with respect to efficient trajectory optimization. These two parameters are responsible for balancing the exploration and exploitation phase that escapes the local optimum. Moreover, $\alpha$ is the ability of each UAV to interact with other UAV in swarm, $\beta$ is the influence of other UAVs on UAVs in neighboring cluster q and is equal to 1, whereas, $\omega$ is a constant. $X_{p, q}$ is the fitness of UAV in the cluster, whereas, $X_{\text {best }}$ is the fitness of best UAV in the entire swarm. F refers to Flag that provides diversity by changing the direction of UAV in the search space.
\begin{equation}\label{eq6}
\begin{aligned}
&U\_{w}=N (\operatorname{rand}(a\_{2}-a\_{1})+a\_{1}), \\
&a\_{1}=0.1 \text { and } a\_{2}=0.2
\end{aligned}
\end{equation}
where $U_{w}$ indicates the worst UAVs and N is the number of search UAVs.

\begin{equation}\label{eq7}
G\_{(p, q)}=G\_{\min (p, q)}+r (G\_{\max (p, q)}-G\_{\min (p, q)})
\end{equation}
where, $G_{(p, q)}$ refers to the position of worst UAV \text{and} $G_{\max (p, q)}, G_{\min (p, q)}$ are the bounds.

Eventually, Algorithm 1 shows the pseudocode for HGO algorithm, including UAV population initialization, clustering and finding the best UAV haivng global best cost.

\SetKwInput{KwInput}{Input}                
\SetKwInput{KwOutput}{Output}              

\begin{algorithm}[]
\DontPrintSemicolon
  
  \KwInput{Initialize UAV population $U_p(p = 1, 2, . . . N)$}
  \KwData{Input Parameters $H_q,P_p,q, C_q, O_1, O_2, O_3$}
Divide the UAV swarm population into cluster sets having same Henry coefficient ($Hq$)\;
Analyse every cluster q in the search space\;
Get the best UAV swarm($U_{p,best}$ ) in every cluster and the best global cost for UAV swarm($U_{best}$)\;

\While{$t$$<$$iter\_{max}$} {
 
    \For{each UAV swarm }
    {
    	 Update the positions of each UAV using Eq. (\ref{eq5})\;

    }
    
Update Henry’s coefficient of each UAV using Eq. (\ref{eq3})\;
Update the trajectory for each UAV to reach the affected area using Eq. (\ref{eq4})\;
Escape from local optima by selecting and ranking the worst UAVs in search space using Eq. (\ref{eq6})\;
Update the position of the worst UAV trajectory using Eq. (\ref{eq7})\;
Update $U_{p,best}$ , and $U_{best}$\;
}

$t=t+1$\;
return $U_{best}$\;
\caption{Pseudo code of HGO algorithm}
\label{al1}
\end{algorithm}

\subsection{Exploration and Exploitation Phase}
The exploration and exploitation phase are balanced by fine tuning the randomness between these phases, allowing the algorithm to jump from local optimum towards globally exploring the search space. The three main control parameters used in the proposed algorithm HGO are $T_{p,q}$, $\alpha$ \text{and} $F$. (a) $T_{p,q}$ refers to the trajectory of UAV p in the cluster q and depends of the time iterations. The best balance between exploration and exploitation is achieved when the search agents (UAV) transfers from global to local phase in search of best UAV. (b) $\alpha$ is the interaction ability of UAV with neighboring UAV in cluster q and aims to transfer the best among them from global to local and vice versa as per the iterations. (c) $F$ is the flag parameter that allows the search agents to change direction, providing diversity and exploring the search space carefully. 

The exploration and exploitation phase comprise of dimension wise diversity measurement and are discussed in this paper. Intuitively, the increased mean value between the search agents refers to the exploration phase. While, in exploitation phase, the reduced mean value between the search agent refers to their closeness to each other. The insignificance difference between the mean diversity value refers to the optimal convergence state of the algorithm. The percentage measurement for exploration and exploitation phase are as follows:

\begin{equation}
\begin{aligned}
&\text { Exploration}=(\frac{\operatorname{D}^{t}}{\operatorname{D}\_{\max }} \times 100)\% \\
&\text { Exploitation}=(\frac{|\operatorname{D}^{t}-\operatorname{D}\_{\max }|}{\operatorname{D}\_{\max }} \times 100)\%\\
\end{aligned}
\end{equation}
where, $\quad \operatorname{D}^{t}=\frac{1}{N} \sum_{q=1}^{D} \operatorname{D}_{q} and \operatorname{D}_{q}=\frac{1}{N} \sum_{p=1}^{N} \operatorname{median}(x^{q})-x_{p}^{q} $. $D^t$ refers to the population diversity value of search agent, $D_{max}$ is the maximum diversity in all iterations, N refers to the size of population, $x_p^q$ indicates the $p^{th}$ population individual having $q^{th}$ dimension in the search space.

\subsection{Pseudo Code}
~Algorithm~\ref{al1} explains the pseudo code of HGO algorithm. The UAV population and input parameters are initialized for trajectory optimization. First the UAV population is divided into clusters that form swarm having same Henry coefficient. Then, these clusters are analyzed and evaluated in the search space. In the next step, based on the evaluation, we get the best UAV swarm and best global cost for UAV swarm. The UAV position are updated in the swarm till it reach the maximum number of iterations. Meanwhile, the optimized trajectory is calculated for each UAV while the Henry coefficient is updated to reach the target location having minimum transportation cost and computational time. The worst UAVs are updated based on the trajectory to escape from local optima and their positions are updated. Finally, the best UAV swarm and best global cost for UAV is updated that provides the minimal transportation cost and computational time.

\section{Methodology}
The workspace and problem statement are discussed in this section for UAV trajectory optimization.

\subsection{Workspace Representation}
The main scenario for UAV trajectory optimization is confined in a workspace that comprises of the start and destination points, control points and obstacles as shown in Figure \ref{Figure3}. The algorithm is designed to provide an obstacle free environment for UAVs to reach the destination. 

The control points $c_1$, $c_2$, $c_3$ detects the nearby obstacles in its vicinity and avoid colliding with it to provide a smooth path for UAV movement. The circular obstacles are placed in the search space that acts as hurdles in the movement of UAVs. These obstacles are scattered across the X, Y coordinates, having different dimensions. The position and size of obstacles are determined based on four different scenarios as discussed in Section \ref{section5}. The total path length of UAV trajectory is calculate based on the euclidean distance between start and destination points.

\subsection{Problem Statement}
UAV autonomous trajectory optimization is a complex optimization problem that tends to minimize the distance between start and destination points. For that purpose, we find the global best solution retrieved from the cost function f(x) in a confined search space. 

\begin{figure}[t]
\center
\includegraphics[width=3in, height = 5cm]{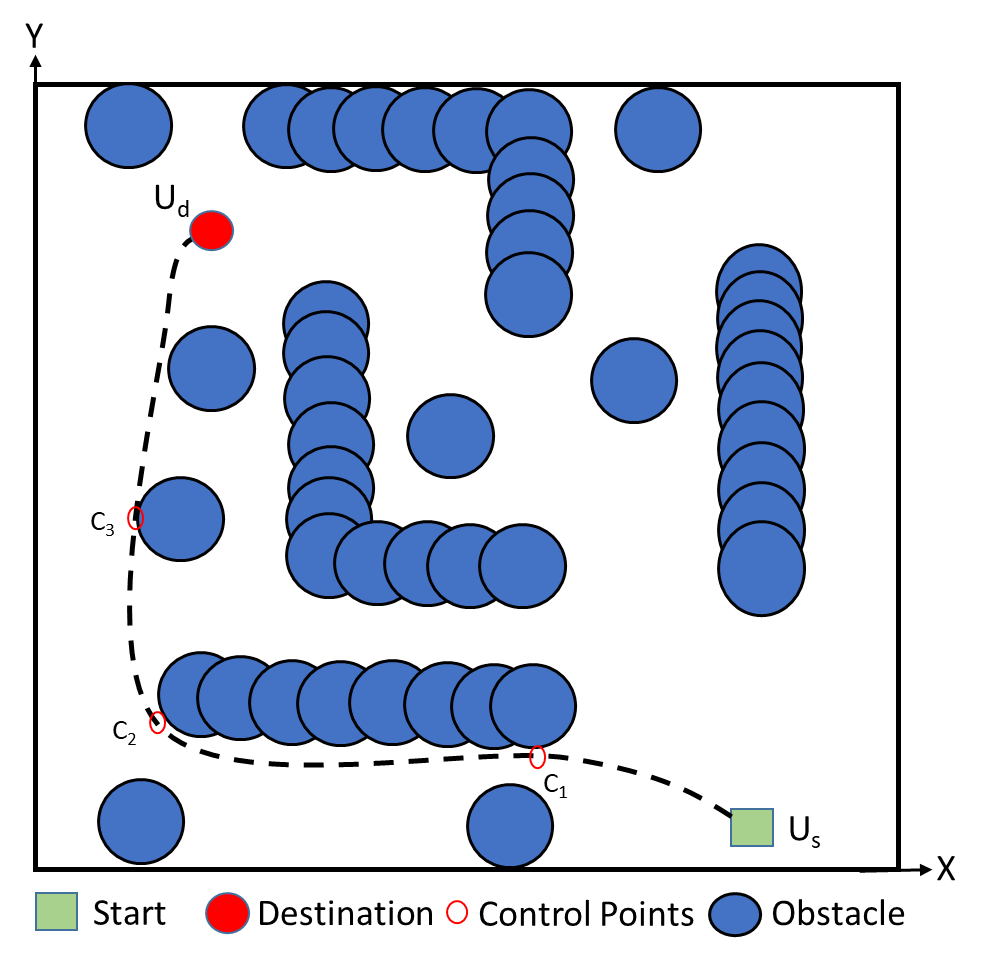} 
\caption{Representation of UAV Trajectory Optimization Workspace.}
\label{Figure3}
\vspace{-4 mm}
\end{figure}

Keeping in mind the optimization problem, collision avoidance plays a significant role in finding the best trajectory for UAVs. The first step towards implementation is considering the control points between the trajectory that are connected to its predecessor and successor in determining the best path as shown in Figure \ref{Figure3}. The control points considered in this study are three $c_1$, $c_2$, and $c_3$. Specifically, the control points ($c_1$, $c_2$, and $c_3$) represent specific positions in the trajectory where adjustments can be made. These points are connected to their predecessors and successors, which means they are part of a sequence that forms the entire trajectory. The primary role of these control points is to define the path's shape and allow for fine-tuning to optimize the trajectory, such as minimizing the distance traveled or avoiding obstacles. The selection of these control points considers the need to balance minimizing the distance traveled with avoiding potential obstacles in the environment. They serve as pivotal locations where the UAV can make adjustments to its path to optimize the trajectory according to the cost.

As we tune these control points, the second important point is to minimize the cost function f(x) for trajectory optimization. The cost function is the Euclidean distance between the start and destination and formulated as follows:   

\begin{equation}
.f(x)=\|U\_{s}-c\_1\|+\sum\_{i=0}^{n=2}\|P\_L-U\_{d}+1\|+\|U\_n+U\_d\|
\end{equation}

\begin{figure}[t]
\center
\includegraphics[width=2.8in, height = 4.5cm]{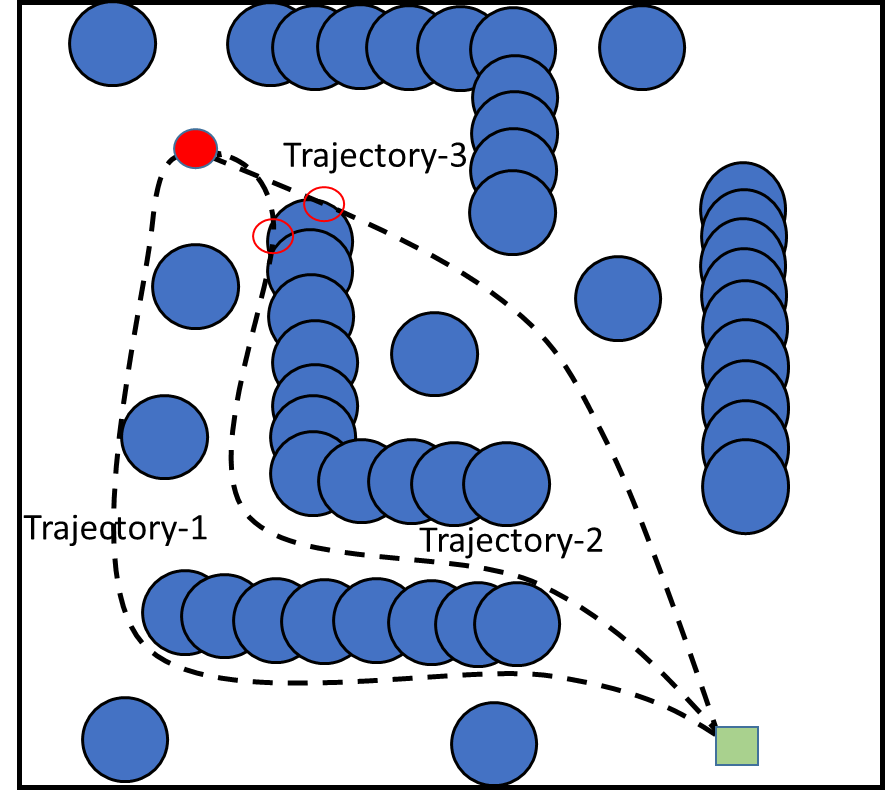} 
\vspace{2 mm}
\caption{Multiple trajectory optimization for UAV from start to destination.}
\label{Figure4}
\vspace{-4 mm}
\end{figure}
\begin{figure}[t]
\center
\includegraphics[width=3.2in, height = 7.5cm]{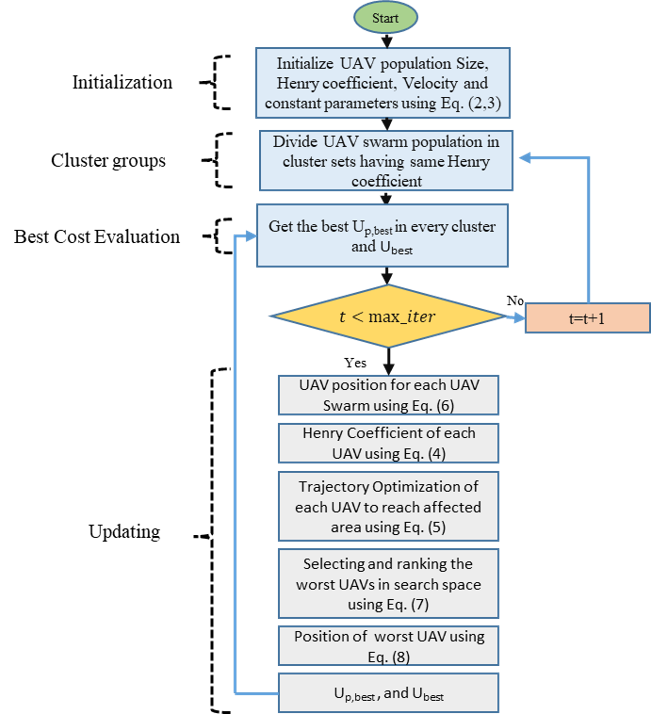} 
\vspace{2 mm}
\caption{Flow chart using HGO algorithm for UAV Trajectory optimization.}
\label{Figure5}
\vspace{-4 mm}
\end{figure}

\subsection{HGO-based Trajectory Optimization}

UAV trajectory optimization based on the HGO algorithm randomly initializes the control points between the start and destination in the search space. As the cost function is calculated, the position of control points is updated automatically. These control points ensure the trajectory is smooth enough and avoid any obstacle in its path. An obstacle violation flag is high as the UAV comes into the vicinity of any obstacle or tries to pass by. Thus, it provides an obstacle-free environment for UAVs for post-disaster assessment like bushfire regions.

Figure \ref{Figure4} shows multiple trajectories for UAVs from stat to destination points. As shown, Trajectory-3 is the shortest but collides with the obstacle in the workspace. Similarly, Trajectory-2 also collides with obstacles, thus causing UAV to be damaged and not complete its task time efficiently. Though, Trajectory-1 is the longest but provides a collision-free path for UAVs to reach the destination points. There are also control points in Figure \ref{Figure4}. In detail, the control points are crucial in determining the feasibility and safety of different trajectories. The figure illustrates multiple trajectories, each associated with a different set of control points. Trajectories-2 and 3, for example, have control points that lead to collisions with obstacles, while Trajectory-1, despite being the longest, has control points that ensure a safe, collision-free path. The selection of control points in Figure 4 highlights the trade-offs between path length and safety, with control points chosen to prioritize either efficiency (shorter distance) or safety (avoiding obstacles).

The flow chart of UAV trajectory optimization using the HGO algorithm is shown in Figure \ref{Figure5}. Where the UAV population and input parameters are initialized, the UAV's best cost is evaluated from the cluster groups based on the trajectory (transportation cost and computational time). The parameters are updated on each iteration till the maximum iteration is reached and the best global cost is calculated. Specifically, at first, the population size, Henry coefficient, velocity and several constant parameters are initialized. Then, we divide UAV swarm population in cluster sets owning same Henry coefficient. After getting the best $U_{p,best}$ in every cluster and $U_{best}$, the judgment execution starts. If $t$ is larger than $max\_iter$, the process of cluster grouping and best cost evaluation repeats. Otherwise, updating process begins, which includes the updating of UAV position, Henry coefficient, trajectory, information of worst UAVs, $U_{p,best}$ and $U_{best}$.


\begin{table*}[]
\centering
\caption{Comparison of Metaheuristic algorithms for UAV Trajectory Optimization}
\label{table 2}
\resizebox{\textwidth}{!}{%
\begin{tabular}{cccccccccc}
\toprule
  \multirow{2}{*}{Algorithm} &
  \multicolumn{2}{c}{Case 1} &
  \multicolumn{2}{c}{Case 2} &
  \multicolumn{2}{c}{Case 3} &
  \multicolumn{2}{c}{Case 4} \\ \cmidrule(l){2-9} 
 &
  Cost (km) &
  Algorithm Complexity (s) &
  Cost (km) &
  Algorithm Complexity (s) &
  Cost (km) (s) &
  Algorithm Complexity (s) &
  Cost (km) (s) &
  Algorithm Complexity (s) \\ \midrule
HGO	& 7.535	& 75.84 &	8.266 &	101.23 & 12.92 &	121.44 &	13.153 &	125.66  \\ \midrule
GWO	& 8.554	& 78.84	& 8.311	& 104.33	& 13.23	& 125.02	& 13.312 &	130.17 \\ \midrule
BMO	& 9.091	& 79.11	& 8.798	& 106.76	& 13.26	& 124.67	& 14.168	& 141.13  \\ \midrule
CSA	& 10.43	& 78.02	& 9.112	& 124.01	& 13.36	& 129.65	& 15.591 &	136.08 \\ \midrule
PSO	& 12.41	& 91.15	& 12.640	& 132.56	& 13.43	& 132.57	& 14.237	& 129.65  \\ \bottomrule
\end{tabular}%
}
\end{table*}

\begin{figure}
  \centering
  \begin{subfigure}[t]{0.4\textwidth}
    \includegraphics[width=\linewidth, height=12em]{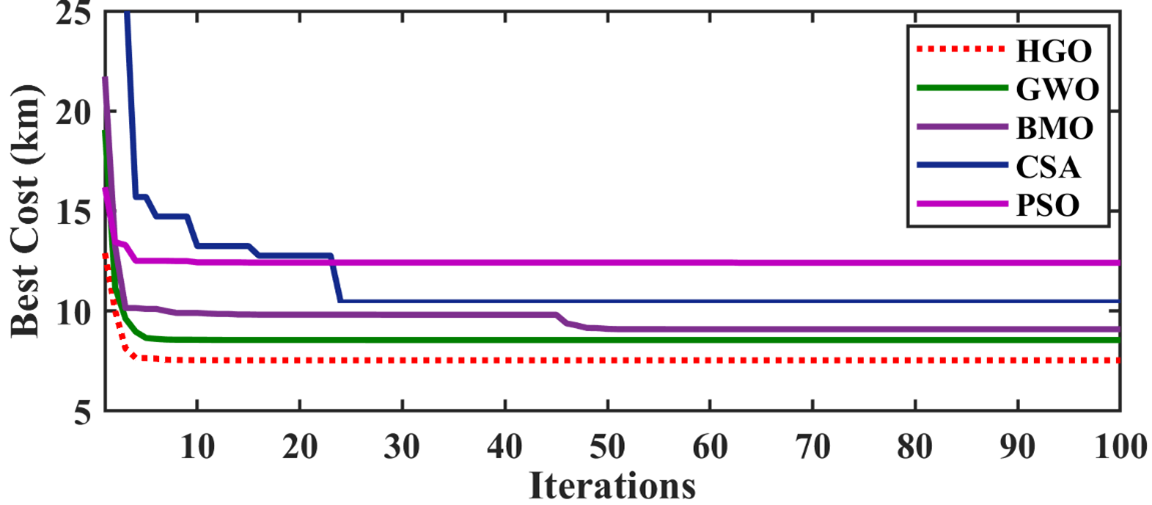}
    \caption{Best Cost Vs Iterations}
    \label{1a}
  \end{subfigure}
  \begin{subfigure}[t]{0.45\textwidth}
    \includegraphics[width=\linewidth, height=14em]{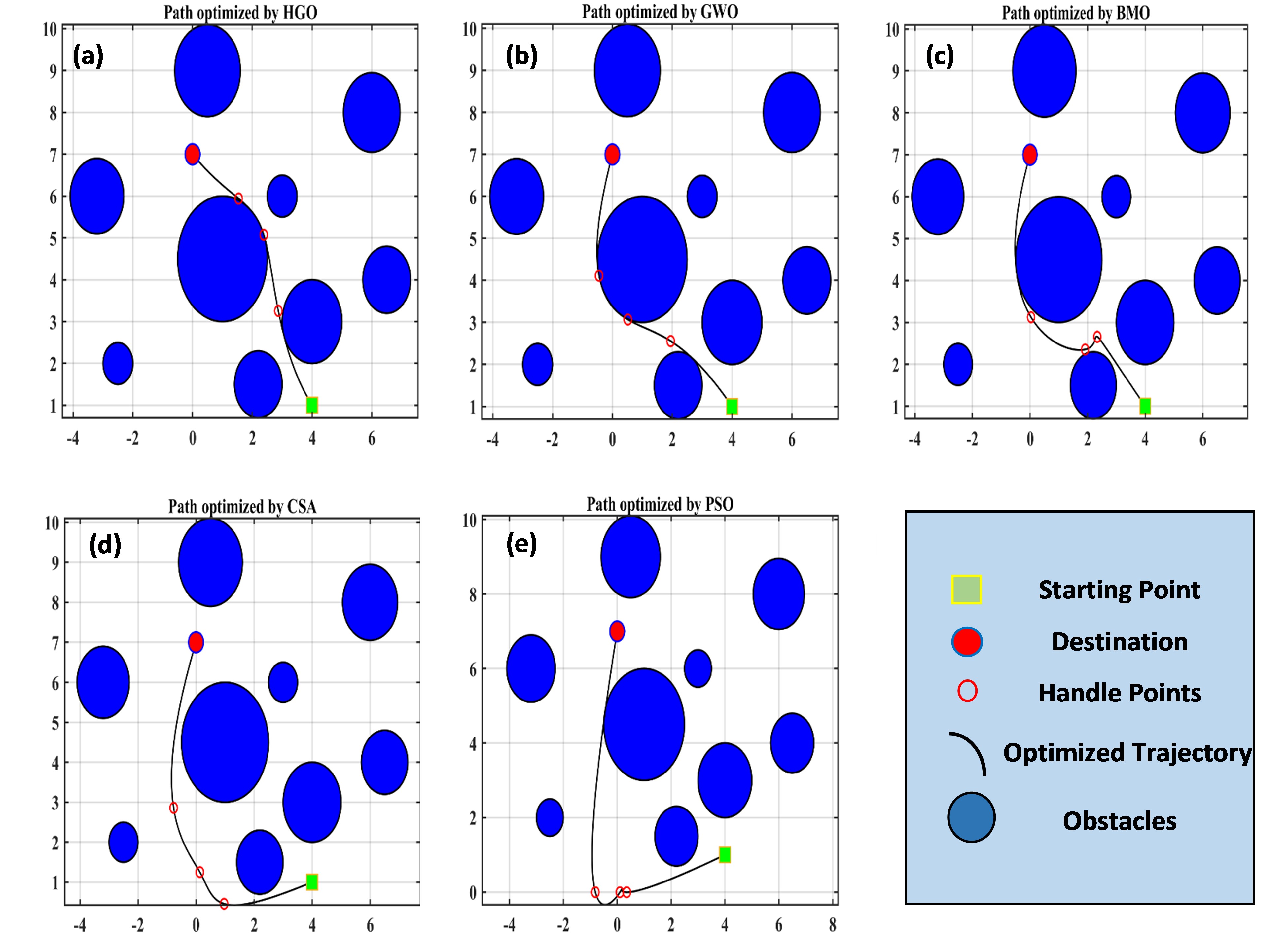}
    \caption{2D Representation of proposed Metaheuristic algorithms}
    \label{1b}
  \end{subfigure}
  \caption{UAV Trajectory Optimization Comparison in Ambient Environment}
  \label{Case1}
\end{figure}

\section{RESULTS AND DISCUSSION}
\label{section5}
In this section, four  of  the  dynamic  environments  such  as,  ambient environment,constrict  environment,  tangle  environment and  complex  environment  have  been  incorporated. The real-time experimental results of proposed algorithm HGO is compared with the state-of-the-art four different metaheuristic algorithms GWO, BMO, CSA and PSO. The simulation environment comprise of following hardware specifications: Intel Core i7-8700 CPU with 16GB RAM using MATLAB 2021a. The comparison parameters included consist of transportation cost and algorithm complexity for finding the optimized trajectory. The results elaborates the performance efficacy of HGO algorithm in different cases.
\begin{itemize}
\item GWO \cite{gwom,gwo3}: This is a heuristic algorithm that mimics the hunting behavior of grey wolves, utilizing the roles of alpha, beta, delta, and omega wolves to find optimal solutions.
\item BMO \cite{bmom}: BMO is a heuristic optimization algorithm inspired by the reproductive behavior of barnacles. It simulates the mating strategies of barnacles in a marine environment, using mechanisms such as mating, reproduction, and adaptive competition among individuals to explore the solution space. 
\item CSA \cite{csam,csa2}: CSA is a heuristic optimization method inspired by cuckoo birds' brood parasitism. It uses Levy flights to generate new solutions and replaces the worst solutions while retaining the best ones, enhancing search efficiency for complex optimization problems.
\item PSO \cite{psom,pso4}: PSO is an optimization algorithm inspired by the social behaviors of birds. It employs a swarm of "particles" that explore the solution space, with each particle adjusting its position based on personal experience and the swarm’s overall performance.
\end{itemize}

\subsection{Case 1: Ambient Environment}
In Case 1, the ambient environment comprise of number of obstacles having different sizes and are spread across the search space. The main aim to analyze these metaheuristic algorithm is there unique ability to optimize any complex problem. UAV trajectory optimization is of paramount importance in any disaster situation such as  bushfire, earthquake and flood where emergency relief team can evacuate the survivals in minimal possible time. Subsequently, HGO algorithm makes it effective for trajectory optimization while balancing the exploration and exploitation phase and reducing the local optima for finding the best global solution. Figure \ref{Case1} shows the comparison between different metaheuristic algorithms in terms of best cost. HGO algorithm outperform the other algorithm and achieves the best transportation cost. Table \ref{table 2} shows the cost (km) and time (s) achieved by HGO, GWO, BMO, CSA and PSO to be 7.535, 8.554, 9.091, 10.43, 12.41 and 75.84, 78.84, 79.11, 78.02, 91.15 respectively.

\subsection{Case 2: Constrict Environment}
\begin{figure}
  \centering
  \begin{subfigure}[t]{0.4\textwidth}
    \includegraphics[width=\linewidth, height=15em]{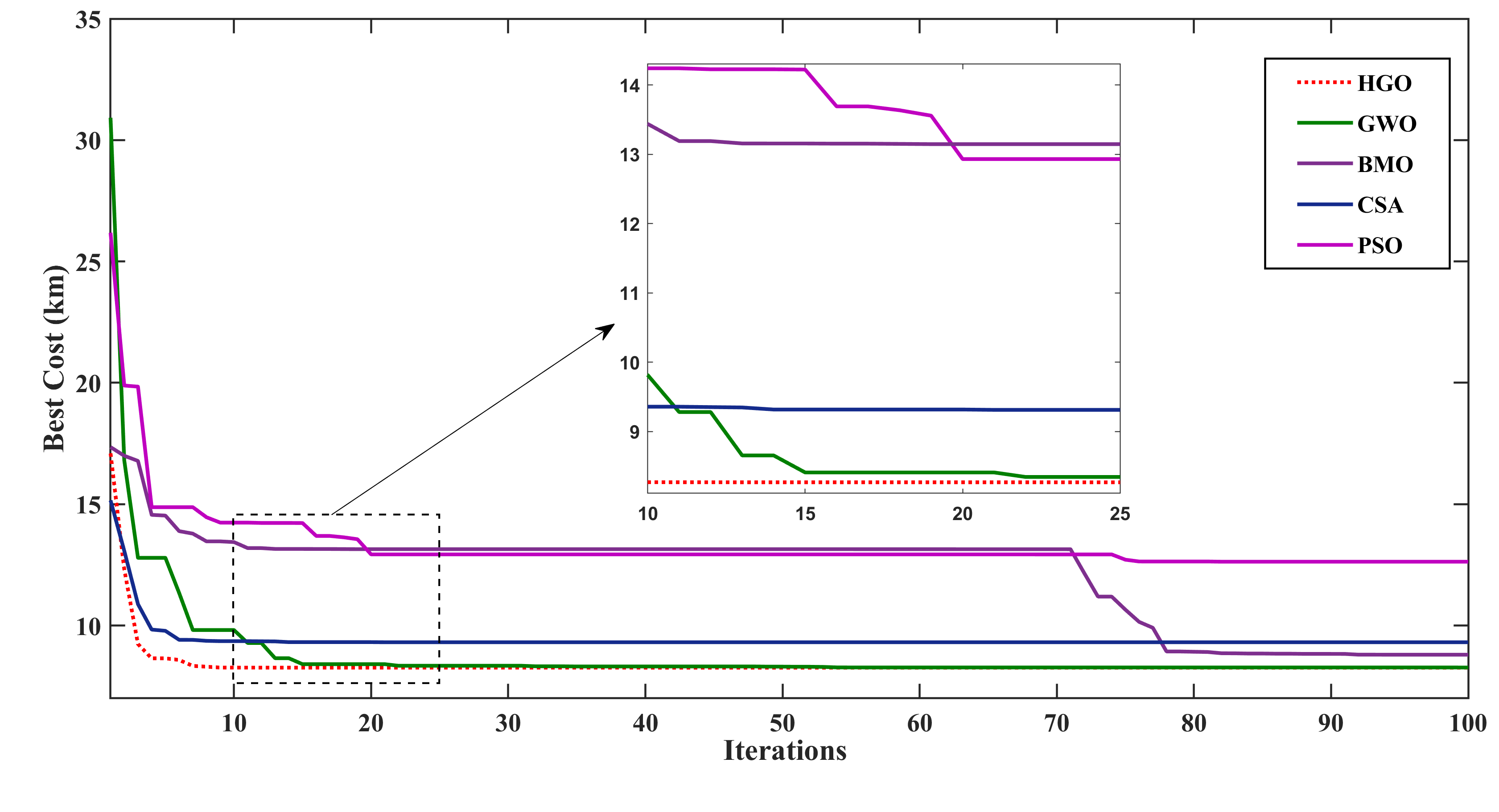}
    \caption{Best Cost Vs Iterations}
    \label{2a}
  \end{subfigure}
  \begin{subfigure}[t]{0.45\textwidth}
    \includegraphics[width=\linewidth, height=15em]{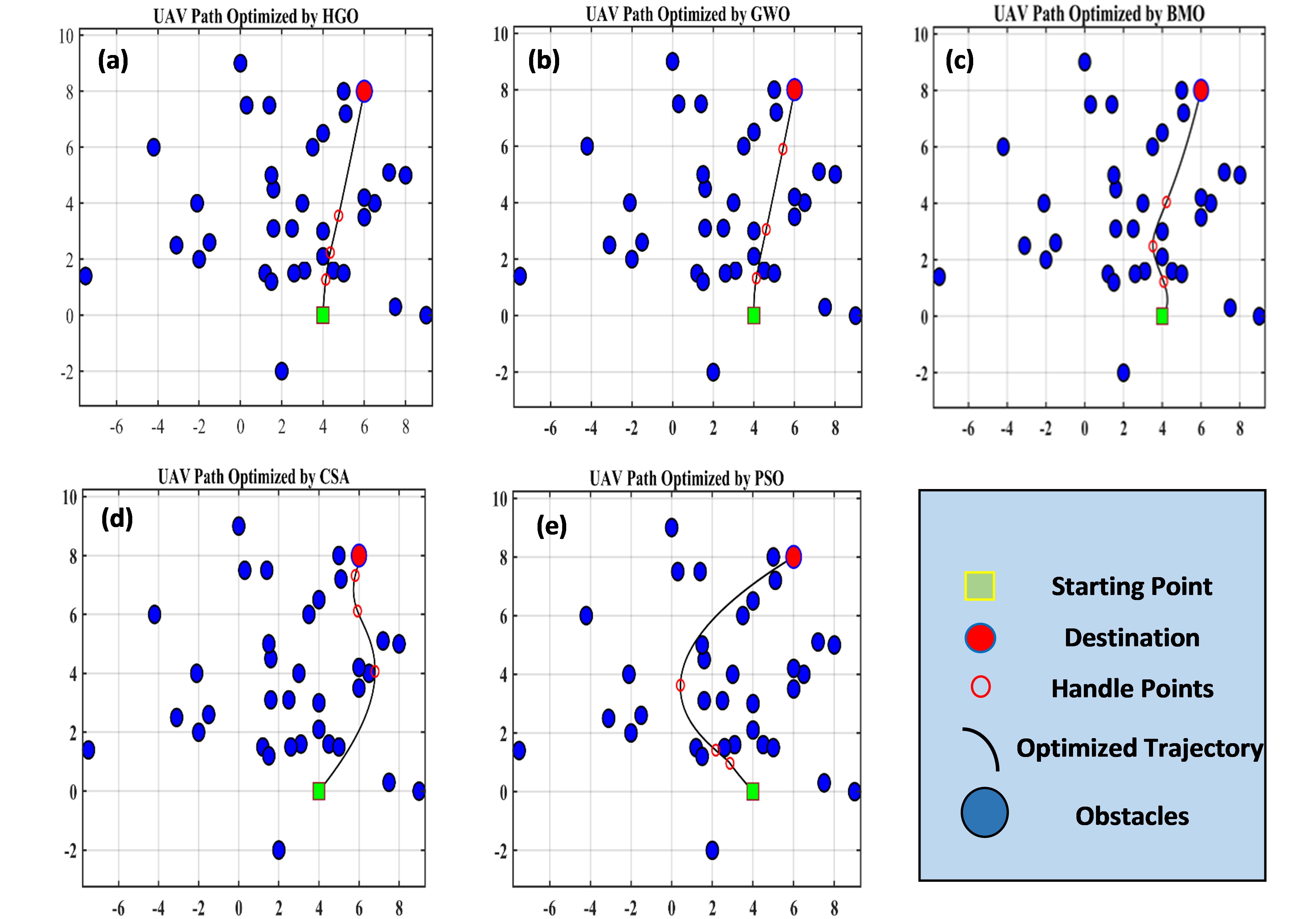}
    \caption{2D Representation of proposed Metaheuristic algorithms}
    \label{2b}
  \end{subfigure}
  \caption{UAV Trajectory Optimization Comparison in Constrict Environment}
  \label{Case2}
\end{figure}

In Case 2, the designed environment is dense that mimics the forest and buildings in the urban as well as rural areas as shown in Figure \ref{Case2}. The effectiveness of the proposed metaheuristic algorithm will be depicted by the transportation cost and time consumed by the UAVs. Obstacles having same radius and different positions along the search space depicts the dense environment where searching for the target location is challenging for UAVs. In this case, PSO algorithm does not perform well and achieves the worst transportation cost of 12.64 km having the highest algorithm complexity of 132.56 s. This trend depicts the poor performance of PSO in dense environment and quickly reaching to the local maxima.

Contrarily, HGO outperforms other algorithms and achieves a best cost of 8.266 (approximately half of PSO) and computation itme of 101.23 s. GWO, BMO and CSA algorithms show moderate results having cost (km) of 8.311, 8.798 and 9.112 respectively. In terms of the computational time, the performance of these metaheuristic algorithms are as follows: HGO $<$ GWO $<$ BMO $<$ CSA $<$ PSO as shown in Table \ref{table 2}.

\subsection{Case 3: Tangle Environment}
\begin{figure}
  \centering
  \begin{subfigure}[t]{0.4\textwidth}
    \includegraphics[width=\linewidth, height=15em]{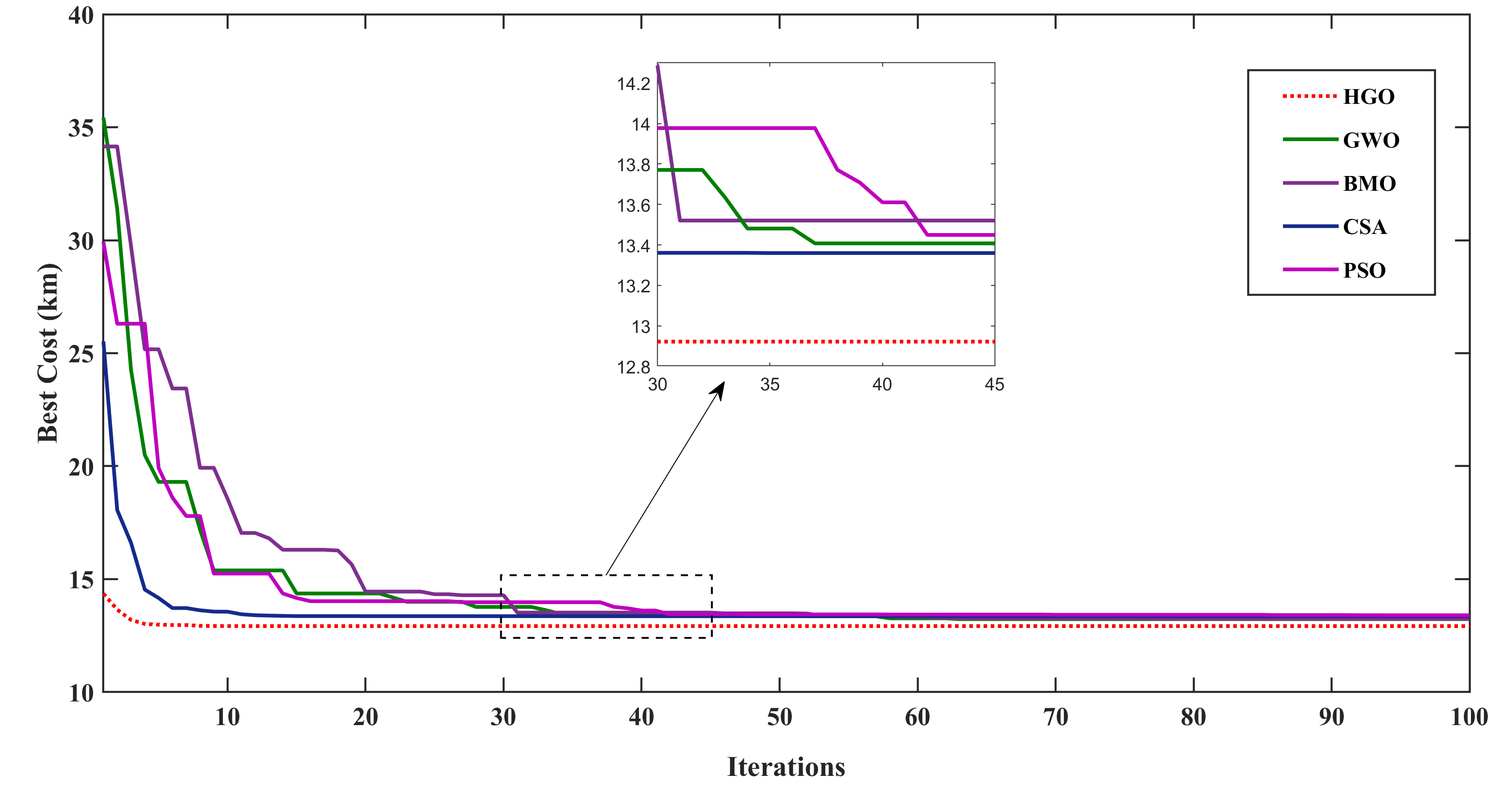}
    \caption{Best Cost Vs Iterations}
    \label{3a}
  \end{subfigure}
  \begin{subfigure}[t]{0.45\textwidth}
    \includegraphics[width=\linewidth, height=13em]{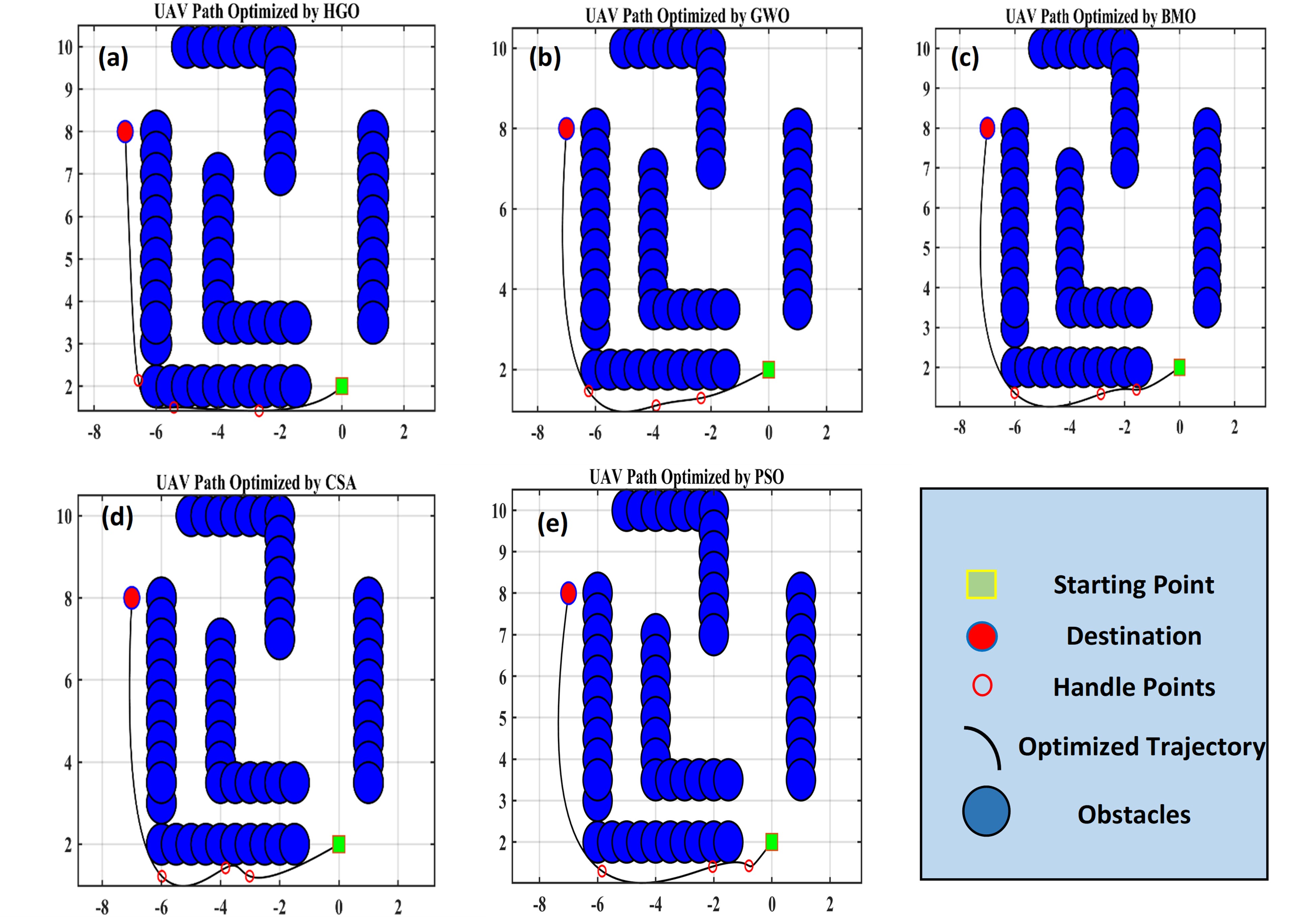}
    \caption{2D Representation of proposed Metaheuristic algorithms}
    \label{3b}
  \end{subfigure}
  \caption{UAV Trajectory Optimization Comparison in Tangle Environment}
  \label{Case3}
\end{figure}

In Case 3, tangle or maze environment is adopted to verify the robustness of the proposed algorithm as shown in Figure\ref{Case3}. The main idea behind designing this environment is that UAV can create a shortest path while swiftly avoiding the hurdles.

In this tangle scenario, all other algorithm except HGO falls into local optima trap. The transportation cost achieved by HGO is 12.92 km, whereas that achieved by GWO, BMO, CSA, PSO are 13.23, 13.26, 13.36, 13.43 respectively. The parameter decrements in GWO and random oscillations in PSO do not allow them to achieve global optimum solution. The algorithm complexity performance for the aforementioned environment is as follows: HGO$<$ BMO$<$ GWO$<$ CSA$<$PSO as shown in Table \ref{table 2}.

\subsection{Case 4: Complex Environment}

In Case 4, the complex environment is a mixture of both Case 2 and Case 3, having a tangle environment along with randomly distributed obstacle. For this scenario, intelligent UAVs with a sense of avoiding obstacles along with path finding is opted. 

The results in Table \ref{table 2} shows that BMO, CSA and PSO stuck in finding the local maxima and are recommended for UAV trajectory optimization in complex environment. 

\begin{figure}
  \centering
  \begin{subfigure}[t]{0.45\textwidth}
    \includegraphics[width=\linewidth, height=15em]{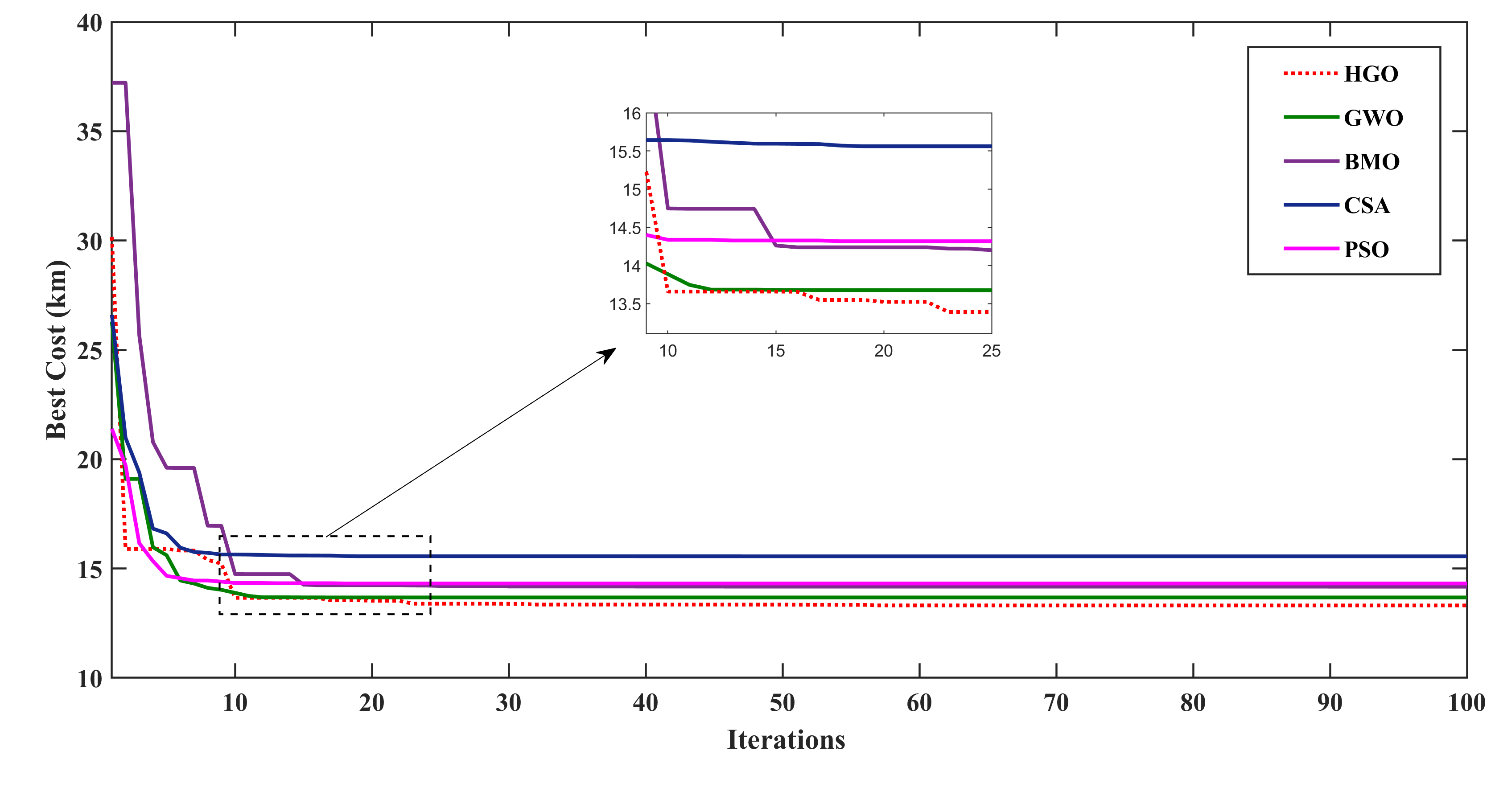}
    \caption{Best Cost Vs Iterations}
    \label{4a}
  \end{subfigure}
  \begin{subfigure}[t]{0.5\textwidth}
    \includegraphics[width=\linewidth, height=13em]{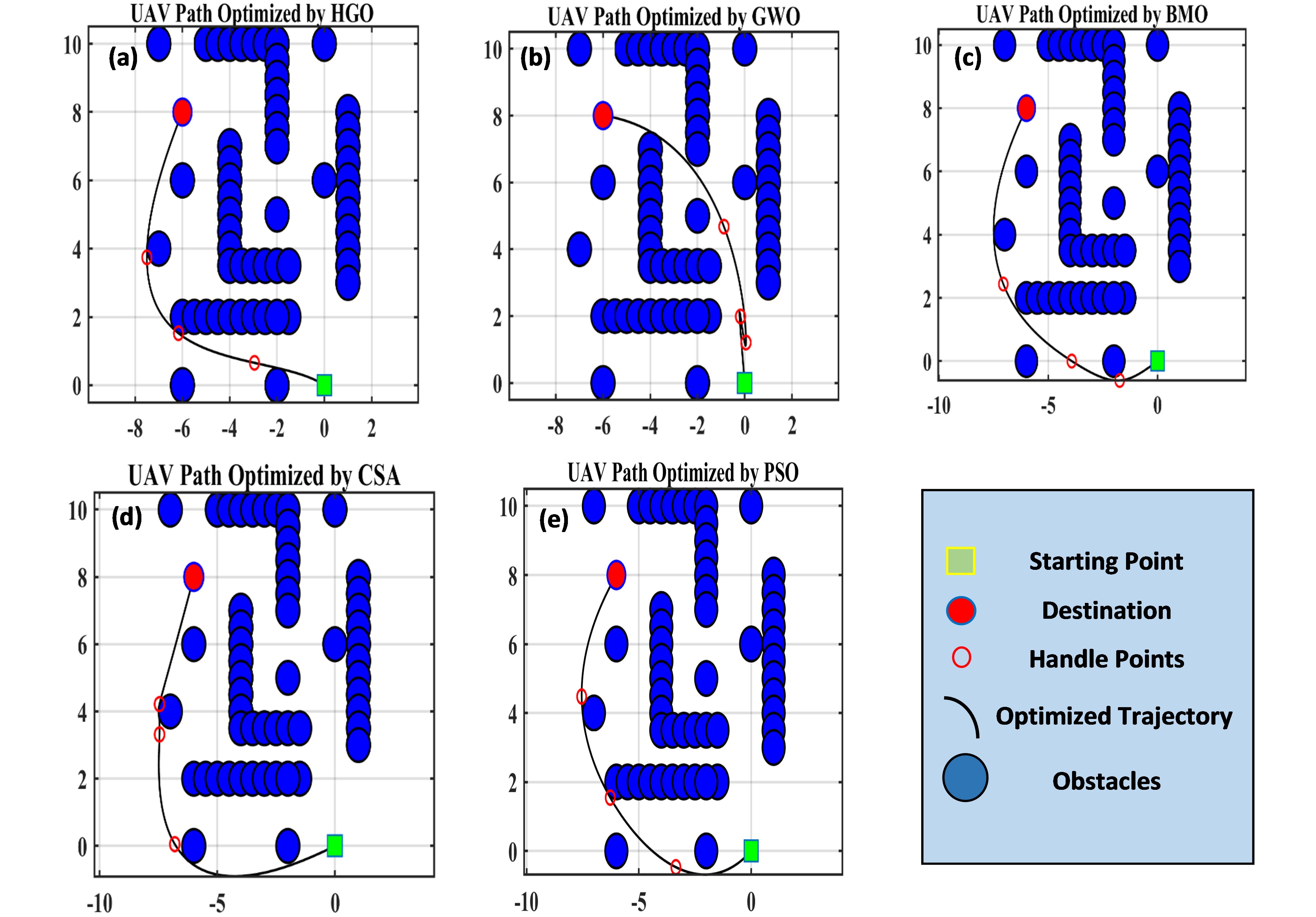}
    \caption{2D Representation of proposed Metaheuristic algorithms}
    \label{4b}
  \end{subfigure}
  \caption{UAV Trajectory Optimization Comparison in Complex Environment}
  \label{Case4}
\end{figure}

Figure \ref{Case4} illustrates that HGO again outperform the other algorithms with a transportation cost of 13.153 km and computational time of 125.66 s. Whereas, the total performance of all the algorithm can be listed as follows: HGO$<$ PSO$<$ GWO$<$ CSA$<$BMO as shown in Table \ref{table 2}.

\subsection{Performance Evaluation}
Experimental results have been carried out to evaluate the trajectory optimization of UAV by leveraging five different state-of-the-art metaheuristic algorithms. The robustness and efficacy of the proposed algorithm is tested based on four different cases that includes ambient environment, constrict environment, tangle environment and complex environment.

The efficient balancing between exploration and exploitation group along with reduced local optima helps in finding the global solution for HGO algorithm in least algorithm complexity. Therefore, HGO algorithm achieves a 39.3\% reduction in transportation cost and 16.8\% reduction in computational time as compared to other metaheuristic algorithms. Therefore, HGO algorithm can be applied to any trajectory optimization problem. The parameters used includes, UAV population size: 100, number of iteration: 100, handle points: 3. 

\section{CONCLUSION AND FUTURE WORK}

In this paper, COS is integrated with efficient trajectory optimization for UAVs in bushfire scenarios and is tested and analyzed using state-of-the-art metaheuristic algorithms. These algorithms including HGO, BMO, GWO, CSA, and PSO are used to solve different types of optimization problems. The experimental results are carried out and verified not only in a static environment but four dynamic environments such as ambient environment, constricted environment, tangle environment, and complex environment. The unique feature of balancing between the exploration and exploitation phases in the HGO algorithm makes the optimization more time efficient, thus reducing the transportation cost for UAV. The results show a significant reduction of 39.3\% in transportation cost and 16.8\% reduction in algorithm complexity for HGO algorithm in an ambient environment.  

The shortcomings of the current work include its inability to address battery consumption issues for extended flight durations. Future work aims to address the battery consumption problem by including the Tethered UAV for long trajectories for bushfire detection where up to 4-5 hours of battery backup is required. Additionally, the current design has relatively weak long-range communication capabilities. LoRaWan communication will be incorporated in our design to have long range communication and the live streaming video can be captured in real time for further monitoring. 




{
\bibliographystyle{IEEEtran}
\bibliography{references}
}

\end{document}